\documentclass[a4paper, amsfonts, amssymb, amsmath, reprint, superscriptaddress,showkeys, nofootinbib, twoside]{revtex4-1}

\usepackage[english]{babel}
\usepackage[utf8]{inputenc}
\usepackage[colorinlistoftodos, color=green!40, prependcaption]{todonotes}
\usepackage{amsthm}
\usepackage{mathtools}
\usepackage{physics}
\usepackage{xcolor}
\usepackage{graphicx}
\usepackage[left=23mm,right=13mm,top=35mm,columnsep=15pt]{geometry} 
\usepackage{adjustbox}
\usepackage{placeins}
\usepackage[T1]{fontenc}
\usepackage{lipsum}
\usepackage{csquotes}

\usepackage[utf8]{inputenc} %
\usepackage[T1]{fontenc}    %
\usepackage{url}            %
\usepackage{booktabs}       %
\usepackage{amsfonts}       %
\usepackage{nicefrac}       %
\usepackage{microtype}      %
\usepackage{lipsum}
\usepackage{graphicx}
\usepackage{color, soul}
\usepackage{url}            %
\usepackage{booktabs}       %
\usepackage{amsfonts}       %
\usepackage{nicefrac}       %
\usepackage{microtype}      %
\usepackage{lipsum}
\usepackage{soul}
\usepackage{color}
\usepackage{float}
\usepackage{graphicx}
\usepackage{caption}
\usepackage{subcaption}
\usepackage{titlesec}
\usepackage{multirow}
\usepackage{amsmath}
\usepackage{soul,color}

\usepackage[nodisplayskipstretch]{setspace}
\graphicspath{ {./figures/} }

\usepackage[pdftex, pdftitle={Article}, pdfauthor={Author}]{hyperref} %
\begin{document}
\title{Predicting Lattice Phonon Vibrational Frequencies Using Deep Graph Neural Networks}

\author{Nghia Nguyen}
\affiliation{Department of Computer Science and Engineering, University of South Carolina, Columbia, SC 29208}
\author{Steph-Yves Louis}
\affiliation{Department of Computer Science and Engineering, University of South Carolina, Columbia, SC 29208}
\author{Lai Wei}
\affiliation{Department of Computer Science and Engineering, University of South Carolina, Columbia, SC 29208}
\author{Kamal Choudhary}
\affiliation{Materials Science and Engineering Division, National Institute of Standards and Technology, Gaithersburg, MD, 20899}
\affiliation{Theiss Research, La Jolla, CA, 92037}
\author{Ming Hu} 
\email[Correspondence email: ]{hu@sc.edu}
\affiliation{Department of Computer Mechanical Engineering, University of South Carolina, Columbia, SC 29208}
\author{Jianjun Hu}
\email[Correspondence email: ]{jianjunh@cse.sc.edu}%
\affiliation{Department of Computer Science and Engineering, University of South Carolina, Columbia, SC 29208}

\date{\today} %

\begin{abstract}
Lattice vibration frequencies are related to many important materials properties such as thermal and electrical conductivities and also the superconductivities. However, computational calculation of vibration frequencies using density functional theory (DFT) methods is computationally too demanding for large number of samples in materials screening. Here we propose a deep graph neural network based algorithm for predicting crystal vibration frequencies from crystal structures with high accuracy. Our algorithm addresses the variable dimension of vibration frequency spectrum using the zero padding scheme. Benchmark studies on two datasets with 15000 and 35,552 samples show that the aggregated $R^2$ scores of the prediction reaches 0.554 and 0.724 respectively. Our work demonstrates the capability of deep graph neural networks to learn to predict spectrum properties of crystal structures in addition to phonon density of states (DOS) and electronic DOS in which the output dimension is constant. 

\end{abstract}


\maketitle

\section{Introduction}

Almost all solids, such as crystals, amorphous solids, glasses, and glass-like materials have an ordered, disordered, or hybrid ordered/disordered arrangement of atoms. Due to the thermal fluctuation, all atoms in a solid phase vibrate with respect to their equilibrium positions. The existence of a periodic crystal lattice in solid materials provides a medium for characteristic vibrations. The quantized, collective vibrational modes in solid materials are called phonons. The study of phonons serves an important part in solid-state physics, electronics, photo-electronics, as well as other emerging applications in modern science and technology, as they play an essential role in determining many physical and chemical properties of solids, including the thermal and electrical conductivity of most materials. Lattice vibrations have long been used for explaining sound propagation in solids, thermal transport, elastic, optical properties of materials, and even photo-assisted processes, such as photovoltaics. For instance, there are numerous studies that explore the determinant role of electron-phonon coupling in heat conduction \cite{yang2016nontrivial,YANG2020119481,cheng2017role,tang2020thermal,yang2018distinct,qin2017external,qin2018lone}, superconductivity \cite{PhysRevB.99.104520,yue2018electron,yang2018strong,tanaka2017electron,el2021fermiology}, and photoelectronics \cite{zheng2017phonon,davoody2017unexpectedly,choudhary2021hybrid,chiu2021heat}. The acoustic branch vibration mode softening has been identified as the mechanism of superconducting transition rather than the Fermi surface nesting in platinum diselenide, a type-II Dirac semimetal \cite{cheng2017tunable}. A previous study also illustrates the pivotal role played by electron-phonon coupling in photocurrent generation in photovoltaics \cite{palsgaard2018efficient}. Phonon-assisted up conversion photoluminescence has been experimentally observed for CdSe/CdScore/shell quantum dots \cite{ye2021phonon}, which could be exploited as efficient, stable, and cost-effective emitters in various applications. Therefore, predicting the basic behaviors of lattice vibrations, i.e., the lattice vibrational frequencies, is beneficial towards future design of novel materials with controlled or tailored elastic, thermal, electronic, and photo-electronic properties.

Despite the great importance of predicting vibrational properties of crystalline materials, high fidelity computing of lattice vibrational frequencies using a considerably large dataset is not an easy task. The traditional method to obtain the vibrational frequencies of a lattice is diagonalizing the dynamical matrix of a crystal structure to get its eigen-values(frequencies). Herewith, we restrict all our discussions to the $\Gamma$-point frequency only. The difficulty lies in evaluating the large amount of interatomic force constants (IFCs) of a lattice in a highly efficient and accurate fashion, which is required for obtaining the dynamical matrix associated with the vibrational frequencies. Depending on the symmetry, composition, and structural complexity (such as number of species and their ratio) of the crystal, IFC calculations could be time and resource consuming. In any case, the IFCs calculation can be accomplished by either a quantum-mechanical approach, which can be used to obtain phonon’s dispersion relation and even anharmonicity, or semi-classical treatment of lattice vibrations, which solves the Newton’s law of mechanics with empirical interatomic potentials. However, the quantum-mechanical approach, despite its high accuracy, cannot be used to evaluate or predict the lattice vibrational frequencies of large amount of crystals with diverse compositions and lattice complexity, due to its high-demand and unbearable computation cost. On the other hand, the empirical potential method, although very fast compared to quantum-mechanical approach, fails to give satisfactory results in most of time. For example, if the interatomic interactions are not accurately calculated, the dynamical matrix could be ill defined and as a result there could be negative values in the obtained frequencies. To this end, developing some algorithms that can accurately and quickly screen and evaluate a large number of crystals will be very promising for high-throughput computing and novel materials design.

Big data and deep learning approaches have already brought a transformative revolution in computer vision, autonomous cars, and speech recognition in recent years. Machine learning and deep learning algorithms have been increasingly applied in materials property prediction \cite{pilania2013accelerating,dunn2020benchmarking,louis2020graph,fung2021benchmarking} and materials discovery \cite{xiong2020evaluating,zhao2021high}. It has been well-acknowledged that, machine learning has the potential to accelerate novel materials discovery by predicting materials properties at very low computational cost and maintaining high accuracy sometimes even comparable to first-principles level at the same time. Although most of time training a good machine learning model would require a decent amount of high-quality data, which is usually obtained through high precision ab initio simulations, the machine learning model is very efficient and attractive for screening and predicting large amounts of unexplored structures and data, which is orders of magnitude faster than traditional one-by-one computation. Among all the methods for materials property prediction, the structure-based graph neural networks have demonstrated [23] the best overall performance with big advantage over composition-based methods and heuristic structure feature-based approaches. In the field of lattice vibration (phonon), their potential has yet to be implemented due to the inherent difference between materials data and image/audio data, and lack of sufficient materials data. Since the vibration frequencies of a crystalline material strongly depends on its atomic structure and the structural patterns strongly relevant to this property are not well understood, it is highly expected that the strong learning capability of deep graph neural networks’ representation can be used to train deep learning models for vibrational frequency prediction.

Benefited from 15,000 mixed-type structures and 35,552 rhombohedral structures with $\Gamma$-frequencies that we have recently calculated, this paper presents a new development of graph neural network and deploys the trained neural network model to predict lattice vibrational frequencies of crystal materials. Benchmark studies on these two datasets showed that our DeeperGATGNN model can achieve very good performance with an $R^2$ score of 0.724 when the model is trained and tested with the rhombohedron crystal structures. It also shows good performance when applied to predict cubic crystal structures. The model performance on the smaller dataset with mixed crystal structures is lower with an $R^2$ score of 0.556.
To the best of our knowledge, this is the first work that uses deep (graph) neural network to study phonon frequencies.

\section{METHODS}
\label{sec:headings}
\subsection{Data}

To evaluate the performance of our graph neural network model for vibration frequency prediction, we prepare two datasets. The first dataset is the Rhombohedron dataset which is composed of 35552 rhombohedral crystal structures obtained by DFT relaxation of the generated cubic structures of three prototypes (ABC$_6$, ABC$_6$D$_6$, and ABCD$_6$) by our CubicGAN algorithm, a deep learning based cubic structure generator \cite{zhao2021high}. The second dataset consists of a mixed type of 15,000 crystal structures. For the Rhombohedron dataset, we split it into a training set with 28441 samples and a test set with 7111 samples. For the Mix dataset, we split it into a training set with 12,000 samples and a testing set with 3000 samples. 

\paragraph{Data calculation and collection.}

All the first-principles calculations are carried out using the projector augmented wave (PAW) method as implemented in the Vienna Ab initio Simulation Package (VASP) based on the density functional theory (DFT). The initial crystal structures were taken from the Materials Project database. We then optimized each crystal structure with both the atomic positions and lattice constants fully allowed to relax in spin-unrestricted mode and without any symmetry constraints. The maximal Hellmann-Feynman force component smaller than $10^{-3}$ eV/A, and the total energy convergence tolerance is set to be $10^{-6}$ eV. The Opt-B88vdW functional was taken into account to deal with the long term interactions in the exchange-correlation interaction. The elastic constants were calculated after structure optimization by using the parameter IBRION = 6 in VASP package. The vibrational frequencies at the $\Gamma$-point were extracted from the OUTAR file outputted by VASP.

\paragraph{Constructing training and testing datasets.}
For each crystal structure, we parse its OUTCAR file for vibration frequencies. As some of the vibrational frequencies are imaginary, they would be represented as negative values. Additionally, since each crystal structure has a variable number of atoms, the output has a variable number of vibrational frequencies. Therefore, we first identify the crystal with the largest number of atoms to determine the maximum number of frequencies to predict. For instance, since the crystal with the largest number of atoms in our data set has 14 atoms, it would have 42 vibration frequencies. Then, the output vector dimension is set as 42 for all crystal structures in the dataset, formatted as [1st frequency, 2nd frequency, 3rd frequency, ..., 42nd frequency]. If the number of vibration frequencies is less than 42, the remaining values are padded with zero. 

\subsection{Definition of the vibration frequency prediction problem}

\begin{figure}[htb!] 
    \begin{subfigure}[t]{0.45\textwidth}
        \includegraphics[width=\textwidth]{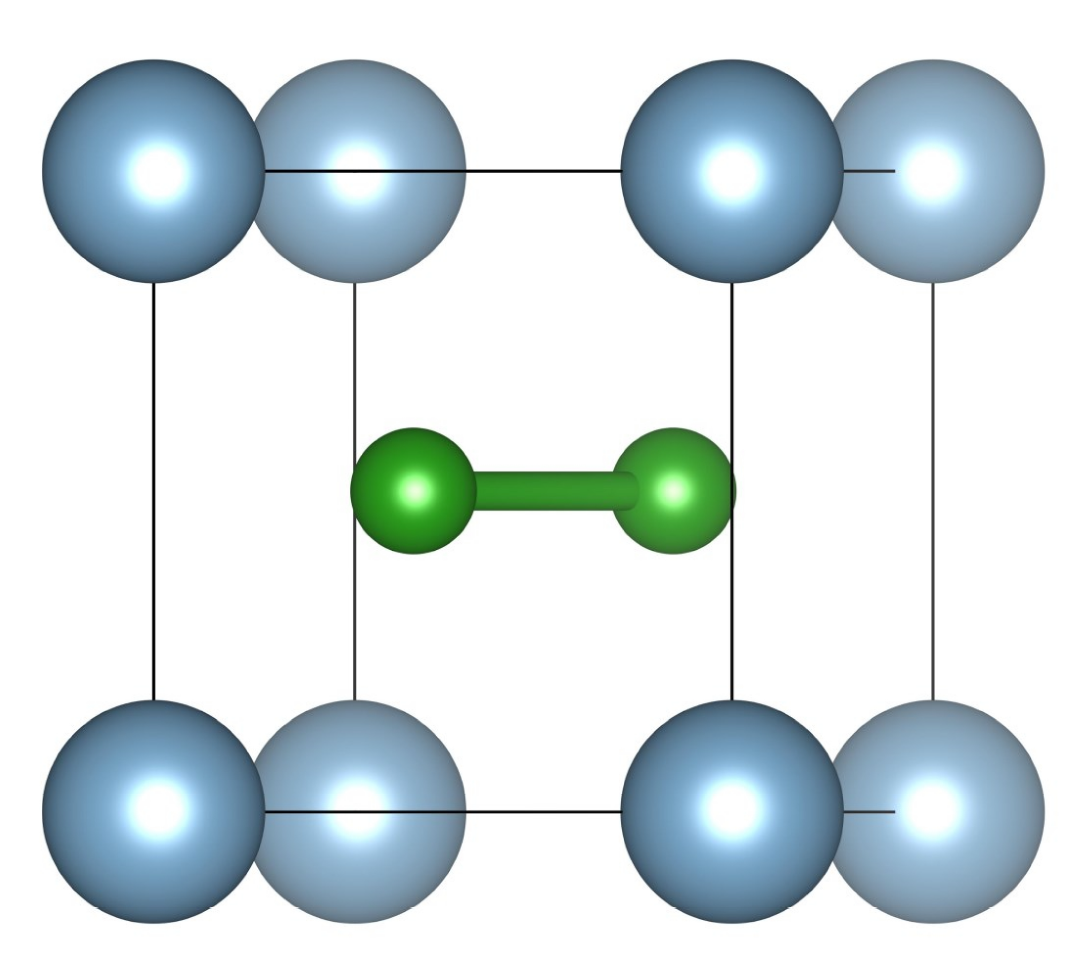}
        \caption{Input graph for compound AlB$_2$}
        \vspace{-3pt}
        \label{fig:Ce2As2O6_predim}
    \end{subfigure}\hfill
    \begin{subfigure}[t]{0.5\textwidth}
        \includegraphics[width=\textwidth]{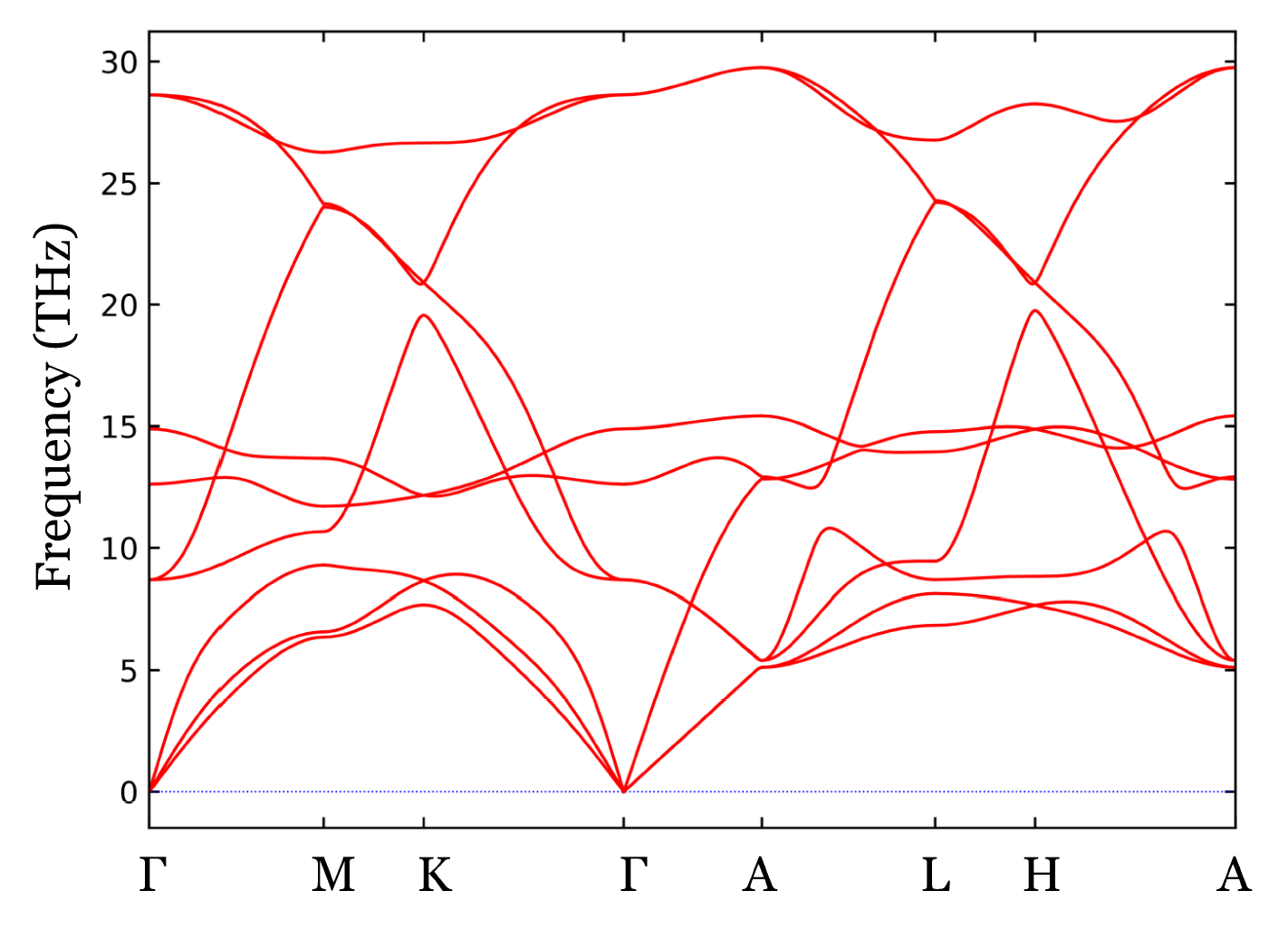}
        \caption{Phonon dispersion of AlB$_2$}
        \vspace{-3pt}
        \label{fig:special_coordinates}
    \end{subfigure}\hfill
    \caption{A representative atomic structure of AlB\textsubscript{2}  (left) and corresponding phonon dispersions (right). The number of phonon frequencies is triple of the number of atoms within the unit cell.}
    \label{fig:regression}
\end{figure}

\paragraph{Task modeling.}
We approach the vibration frequency prediction task as a variable-dimension regression problem (Figure \ref{fig:regression}). For an input POSCAR file, we need to predict its vibration frequencies as a vector of variable dimension. 

\subsection{Scalable Global Attention Graph Neural Network}

\begin{figure*}[tbh!] 
    \begin{subfigure}[t]{0.90\textwidth}
        \includegraphics[width=\textwidth]{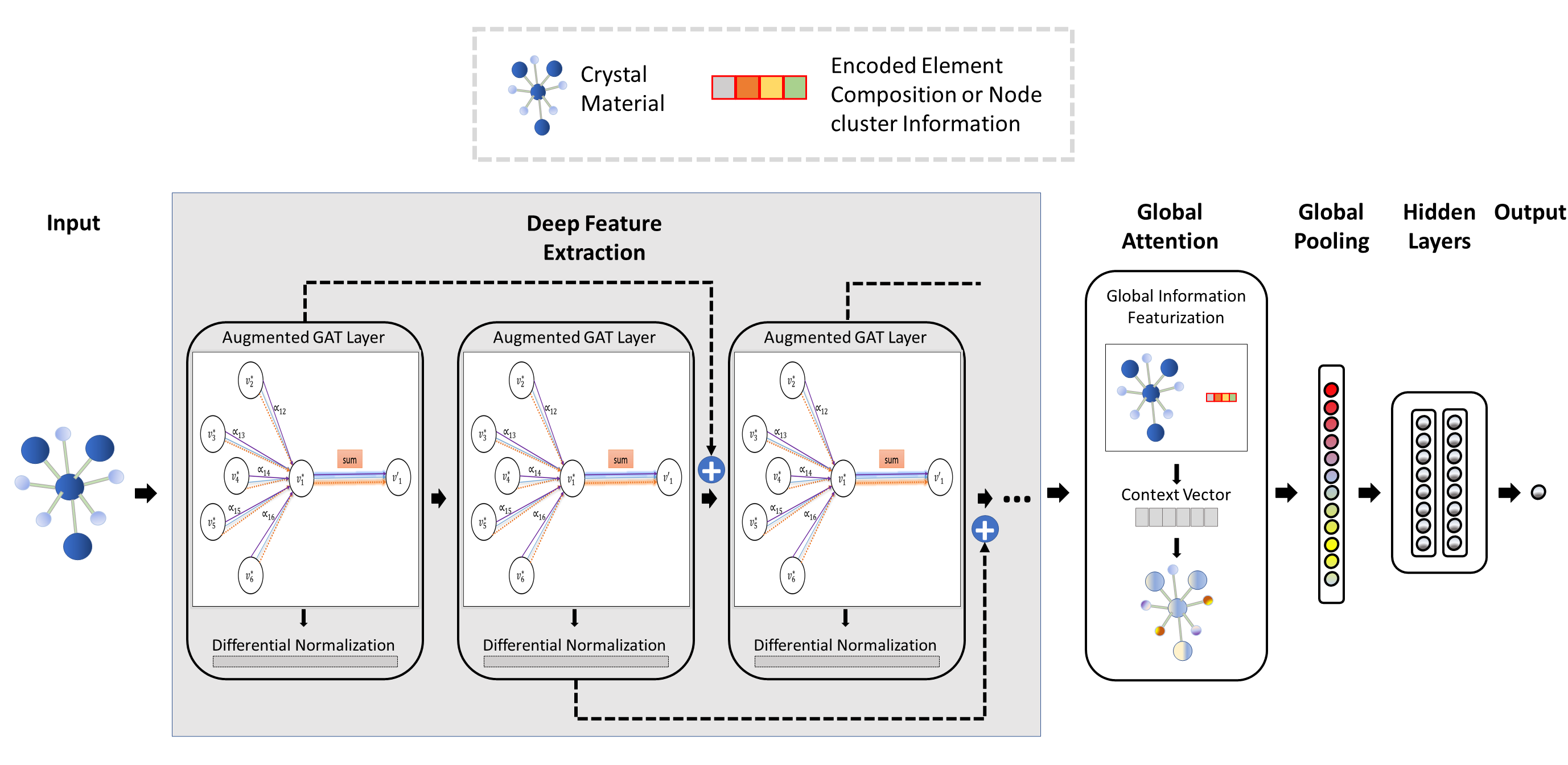}
        \vspace{-3pt}
        \label{fig:Ce2As2O6_predim}
    \end{subfigure}\hfill
    \caption{Architecture of the deeperGATGNN neural network\cite{omee2021scalable}. It is composed of several graph convolution layers with differentiable normalization and skip connections plus a global attention layer and final fully connected layers.}
    \label{fig:framework}
\end{figure*}

To learn the sophisticated structure to property relationship between the crystals and their vibration frequency, we use our recently developed scalable deeper graph neural networks with global attention mechanism \cite{omee2021scalable}. Our DeeperGATGNN model (Figure \ref{fig:framework}) is composed of a set of augmented graph attention layers with ResNet style skip connections and differentiable group normalization to achieve complex deep feature extractions. 
After several such feature transformation steps, a global attention layer is used to aggregate the features at all nodes and a global pooling operator is further used to process the information to generate a latent feature representation for the crystal. This feature is then mapped to the vibration frequencies using a few fully connected layers. To train the model, first we convert all crystal structures of the dataset into graph structures using a radius threshold of 8 Å and the maximum number of neighbor atoms to be 12. The graph representation of our dataset allows us to automatically achieve translation and rotation invariant feature extraction. 

One of the major advantages of our DeeperGATGNN model for materials property prediction lies in its high scalability and state-of-the-art prediction performance as benchmarked over six datasets \cite{omee2021scalable}. The scalability allows us to train very deep network with 10 or more graph attention layers to achieve complex feature extraction without performance degradation as many other graph neural networks suffer due to the over-smoothing issue. Another advantage is that the DeeperGATGNN model has demonstrated good performance without the need of computationally expensive hyper-parameter tuning. The only major parameter is the minimum number of graph attention layers.

\paragraph{Differentible group normalization} One of the key issues of standard graph neural networks is the over-smoothing problem, which leads to the homogenization of the node representation with the stacking of an increasing number of graph convolution layers. To address this issue and build deeper graph neural network, we used a differentiable group normalizer \cite{zhou2020towards} to replace the standard batch normalization. This operator first tries to cluster the nodes based on their representation and then cluster them and do normalization for each cluster.

\paragraph{Residual skip connection} We also added a set of residual skip connections to our GATGNN models, which is a well-known strategy to allow training deeper neural networks as first introduced in ResNet framework \cite{he2016deep} and later used in graph neural networks too \cite{li2021training}. For each of our graph convolution layer, we added one skip connection to it.

\subsection{Evaluation measures}

Our study uses a graph neural network to create a model that predicts vibrational frequency. In order to evaluate its performance, we use mean absolute error (MAE) and the coefficient of determination (R$^2$). Their formulas are as shown below:
\begin{equation}
    MAE = {\frac{1}{n}\sum_{i=1}^{n}|y_{i}-\hat{y}_{i}|}
\end{equation}
\begin{equation}
    R^{2} = {1-\frac{\sum_{i=1}^{n}(y_{i}-\hat{y}_{i})^{2}}{\sum_{i=1}^{n}(y_{i}-\Bar{y})^{2}}}
\end{equation}
where \emph{n} is the number of data points, and $y_i$ and $\hat{y}_i$ are, respectively, the actual and predicted value for the \emph{i}th data point in the data set. The variable $\Bar{y}$ is the mean value of the all the $y_i$ data points. 

\section{Experimental results}
\label{sec:others}

\subsection{Overall performance of vibrational frequency prediction}

\begin{figure}[htb!] 
    \centering
    \begin{subfigure}[t]{0.5\textwidth}
        \includegraphics[width=\textwidth]{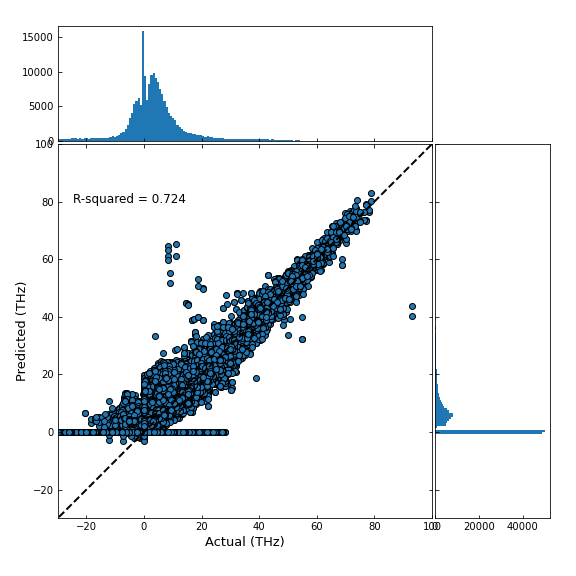}
        \label{fig:special_coordinates}
    \end{subfigure}\hfill
    \caption{Performance of deeperGATGNN for vibration frequencies prediction over the Rhombohedron dataset. The scatter plot shows the predicted versus ground truth vibration frequency for all test materials. }
    \label{fig:performance}
\end{figure}

We first trained a deeperGATGNN model over the more homogeneous structure dataset, the Rhombohedron dataset for vibrational frequency prediction. We randomly picked 28441 samples for training and remaining 7,111 samples for testing. The following hyper-parameters are used for our graph neural network model training: Learning rate = 0.004, graph convolution layers = 10, batch size = 128. No dropout is used as it always deteriorates the prediction performance.
We calculate the MAE for both testing samples and training samples respectively. The average MAE for the training samples is 4.28943 Thz while the average MAE for the testing samples is 4.28879 Thz. To further check the model performance, we show the predicted vibration frequencies versus the ground truth values for all the test samples in the same scatter plot as shown in Figure \ref{fig:performance}. First, we find that most of the points are located around the diagonal indicating a high prediction performance, with its $R^2$ score reaching 0.694. There are a few outliers gathering around low frequency ground truth area. The majority of prediction errors occur for points on the bottom line where a large number of ground truth vibration frequencies are predicted as zero, which maybe due to a systematic deficiency of our model that can be improved. Overall, a majority of vibration frequencies have been predicted with high precision.

\begin{figure*}[htb!] 
    \centering
    \begin{subfigure}[t]{\textwidth}
        \includegraphics[width=\textwidth]{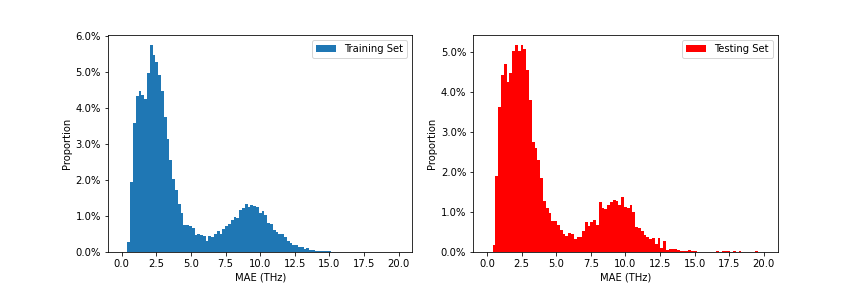}
        \vspace{-3pt}
        \label{fig:Ce2As2O6_predim}
    \end{subfigure}\hfill
     \caption{Histograms of MAE prediction errors over the training samples and the testing samples for the Rhombhedron dataset}
    \label{fig:mae_performance}
\end{figure*}

To check the generalization performance of our deeperGATGNN model for vibration frequency prediction, we plot the histogram of the prediction MAE errors over both the training set and the test set of our Rhombhedron dataset (Figure \ref{fig:mae_performance}). It is found that most frequency MAE errors are around 2.5 Thz while there is another small peak around 9 Thz. It is interesting to find that the MAE histogram over the test set has very similar distribution, indicating the good generalization performance of our model for vibration frequency prediction.

\begin{figure}[h!] 
    \centering
    \begin{subfigure}[t]{0.5\textwidth}
        \includegraphics[width=\textwidth]{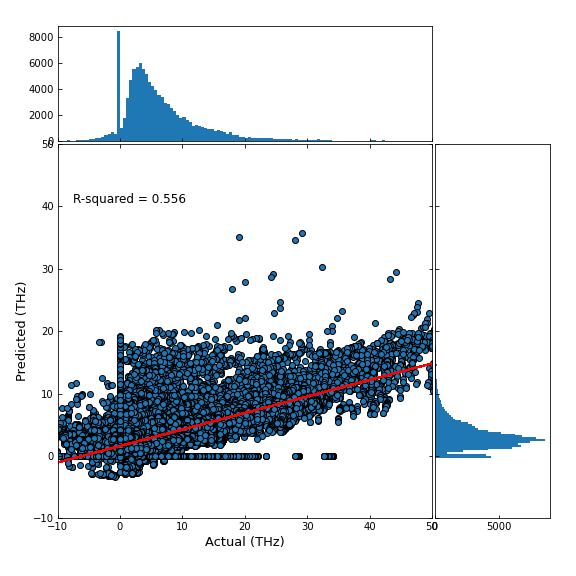}
        \label{fig:special_coordinates_mix}
    \end{subfigure}\hfill
    \caption{Performance of deeperGATGNN for vibration frequencies prediction over the Mix dataset. The scatter plot shows the predicted versus ground truth vibration frequency for all test materials.}
    \label{fig:scatter_mixset_performance}
\end{figure}

In order to further verify the performance of our deeperGATGNN model, we trained the another model using the Mix dataset with more complex and diverse structures compared to the Rhombhedron dataset, which has 15,000 crystal structures. We used a training set with 12,000 samples and a testing set with 3000 samples then calculated the MAE errors and $R^2$ score.  As shown in Figure \ref{fig:scatter_mixset_performance}, the scatter plot of the predicted vibration frequencies versus the ground truth values for all test materials has a much wider distribution around the regression line compared to the result in Figure \ref{fig:performance}. The $R^2$ score here is 0.556, which is significantly lower than 0.694 obtained for the Rhombhedron dataset, indicating the muchg higher challenge in predicting the vibration frequency of mixed structures. Another possible reason is that the Mix dataset has much smaller number of samples: 15,000 versus 35,550. However, we can still see that our deeperGATGNN model has achieved a reasonably good performance overall as shown by the clear trend of the regression line.

\begin{figure*}[htb!] 
    \centering
    \begin{subfigure}[t]{\textwidth}
        \includegraphics[width=\textwidth]{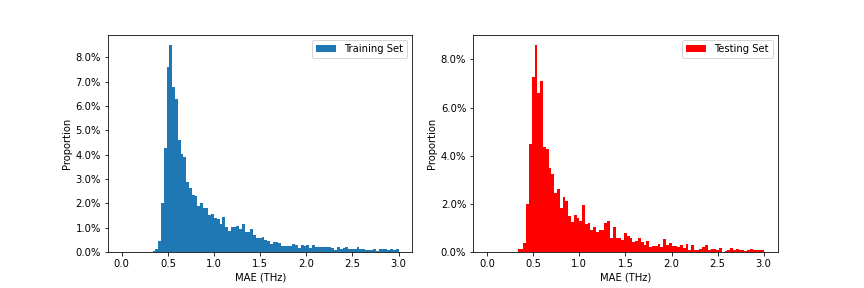}
       
        \vspace{-3pt}
        \label{fig:Ce2As2O6_predim}
    \end{subfigure}\hfill
     \caption{Histograms of the MAE prediction errors of training samples and testing samples for the Mix dataset. }
    \label{fig:mae_mixset_performance}
\end{figure*}

To check the generalization performance of our deeperGATGNN model on the Mix dataset, we show the MAE distributions for both the training set and the testing set in Figure \ref{fig:mae_mixset_performance}. We find that the MAE histograms of the training set and the testing set from the Mix dataset is almost the same, indicating its good generalization performance. An interesting observation is that the MAE distribution for the Mix dataset has only one peak while it has two peaks as shown in Figure \ref{fig:performance}.

\paragraph{Training process and the effect of training set size} \par

\begin{figure*}[th!] 
    \begin{subfigure}[t]{0.45\textwidth}
        \includegraphics[width=\textwidth]{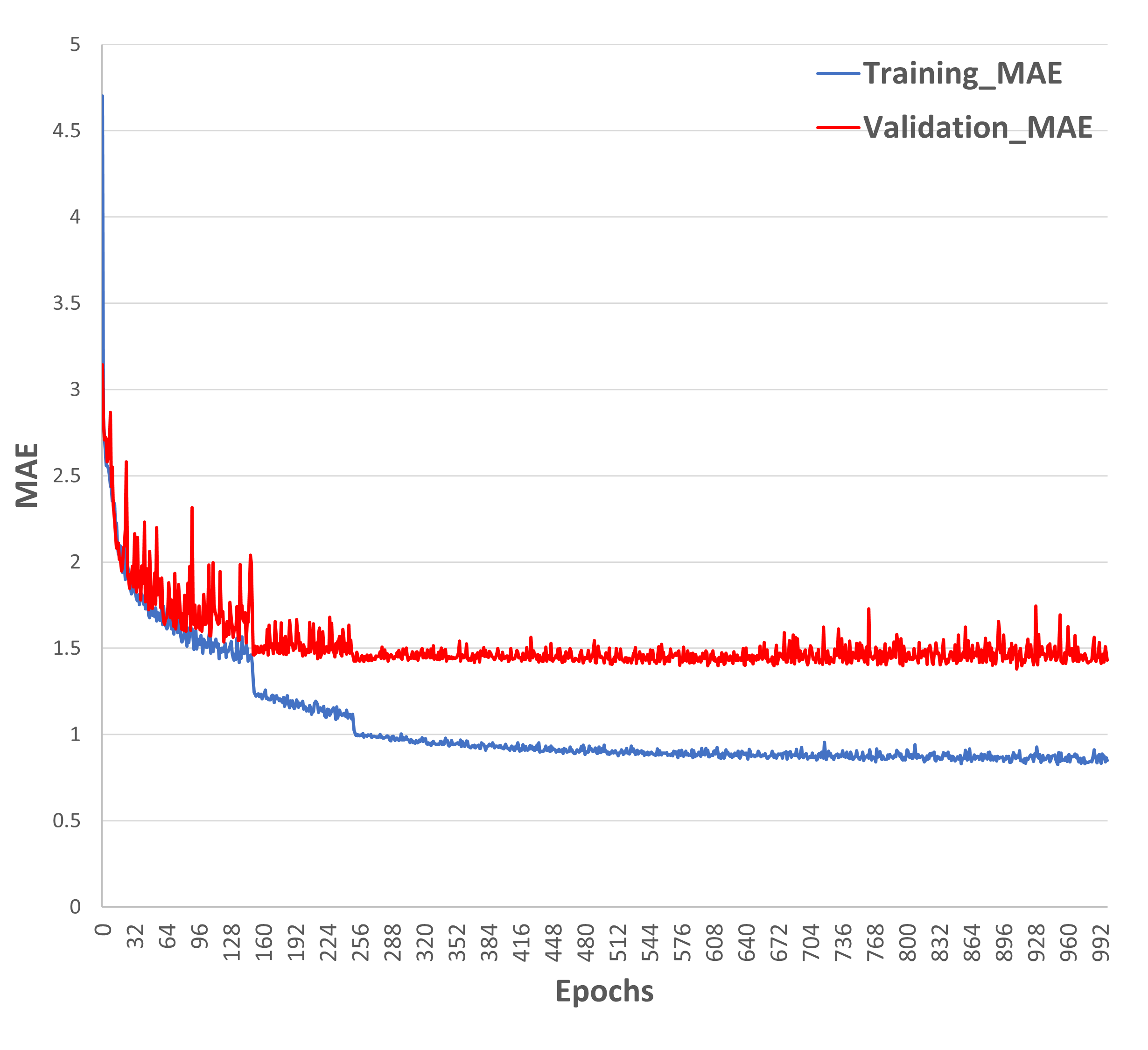}
        \caption{MAE changes during training}
        \vspace{-3pt}
        \label{fig:train_val_error}
    \end{subfigure}\hfill
    \begin{subfigure}[t]{0.45\textwidth}
        \includegraphics[width=\textwidth]{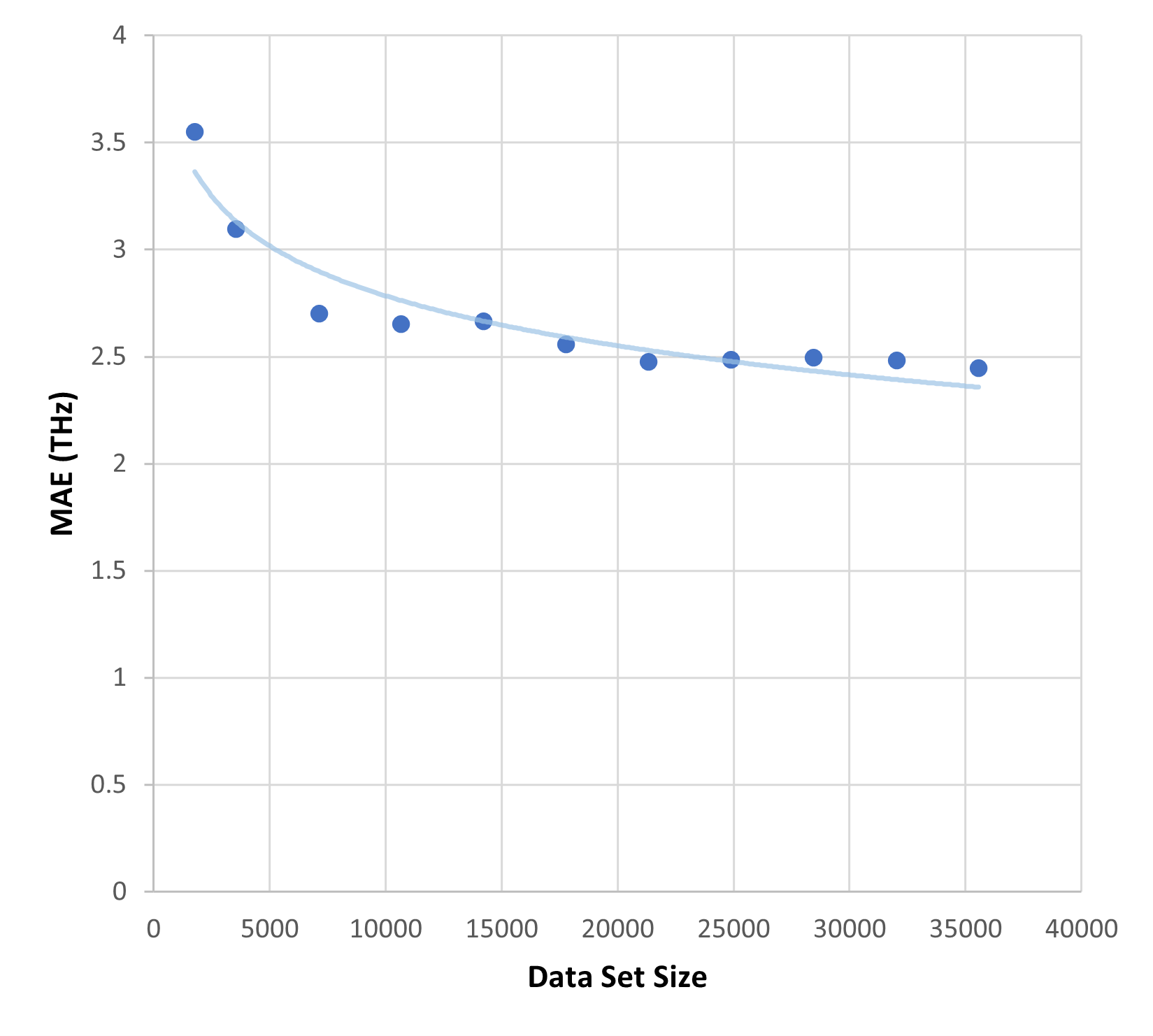}
        \caption{How training set size affects performance}
        \vspace{-3pt}
        \label{fig:performance_train_size}
    \end{subfigure}\hfill
    \caption{The characteristics of the deeperGATGNN model training process.}
    \label{fig:mae_error_train}
\end{figure*}

To understand the model training process of the deeperGATGNN model for vibration frequency, we plotted the training and validation errors during the training process as shown in Figure \ref{fig:train_val_error}. It can be found that the training error keeps going until becoming stagnant while the larger validation errors also go down and become stable after about 300 epochs, indicating the good fitting of the model (no overfitting). We further checked how the training set size may affect the model performance by training different models using different number of training samples of the Rhombhedron dataset. The results are shown in Figure \ref{fig:performance_train_size}. We found that the prediction MAE errors keep going down when more training samples are used. But when the training sample number reaches 20,000, there is no significant performance improvement.

\paragraph{Hyper parameter study:}

It is well known that hyper-parameters of graph neural networks might strongly affect their final performance. To figure out their impact and obtain the optimal settings, we conducted a series of hyper-parameter tuning experiments. The main hyper-parameters of our model include the number of graph convolution layers, the learning rate, the batch size, and the dropout rate (for controlling the overfitting issue). The results are shown in Table \ref{table:hyperparameter1}. First we found that whenever we add the dropout to our model, it leads to worse performance, which is in contrast to the deep neural network models in computer vision. So no dropout is used in our experiments. Second, we find that with a given learning rate ranging from 0.001 to 0.005, the larger batch size (256) usually generates lower performance compared to the result with batch size 128. The optimal performance is obtained with learning rate 0.004, 10 graph convolution (AGAT) layers, and batch size of 128 for all experiments on both data sets. 

\begin{table*}[th!]
\caption{Prediction performance (MAE errors) of different parameter settings}
\label{table:hyperparameter1}
\centering
\resizebox{\textwidth}{!}{\begin{tabular}{c|r|r|r|r|r|r|r|r|r|r|}
\cline{2-11}
\multicolumn{1}{l|}{} & \multicolumn{2}{c|}{\textbf{\begin{tabular}[c]{@{}c@{}}Learning Rate\\ 0.001\end{tabular}}} & \multicolumn{2}{c|}{\textbf{\begin{tabular}[c]{@{}c@{}}Learning Rate\\ 0.002\end{tabular}}} & \multicolumn{2}{c|}{\textbf{\begin{tabular}[c]{@{}c@{}}Learning Rate\\ 0.003\end{tabular}}} & \multicolumn{2}{c|}{\textbf{\begin{tabular}[c]{@{}c@{}}Learning Rate\\ 0.004\end{tabular}}} & \multicolumn{2}{c|}{\textbf{\begin{tabular}[c]{@{}c@{}}Learning Rate\\ 0.005\end{tabular}}} \\ \hline
\multicolumn{1}{|c|}{\textbf{\begin{tabular}[c]{@{}c@{}}AGAT\\ LAYERS\end{tabular}}} & \multicolumn{1}{c|}{\textit{\begin{tabular}[c]{@{}c@{}}Batch Size \\ 128\end{tabular}}} & \multicolumn{1}{c|}{\textit{\begin{tabular}[c]{@{}c@{}}Batch Size\\ 256\end{tabular}}} & \multicolumn{1}{c|}{\textit{\begin{tabular}[c]{@{}c@{}}Batch Size \\ 128\end{tabular}}} & \multicolumn{1}{c|}{\textit{\begin{tabular}[c]{@{}c@{}}Batch Size\\ 256\end{tabular}}} & \multicolumn{1}{c|}{\textit{\begin{tabular}[c]{@{}c@{}}Batch Size\\ 128\end{tabular}}} & \multicolumn{1}{c|}{\textit{\begin{tabular}[c]{@{}c@{}}Batch Size\\ 256\end{tabular}}} & \multicolumn{1}{c|}{\textit{\begin{tabular}[c]{@{}c@{}}Batch Size\\ 128\end{tabular}}} & \multicolumn{1}{c|}{\textit{\begin{tabular}[c]{@{}c@{}}Batch Size \\ 256\end{tabular}}} & \multicolumn{1}{c|}{\textit{\begin{tabular}[c]{@{}c@{}}Batch Size\\ 128\end{tabular}}} & \multicolumn{1}{c|}{\textit{\begin{tabular}[c]{@{}c@{}}Batch Size\\ 256\end{tabular}}} \\ \hline
\multicolumn{1}{|c|}{\textbf{5}} & 1.94823 & 2.33063 & 1.642459 & 1.89346 & 1.67631 & 1.74031 & 1.53829 & 1.52718 & 1.38929 & 1.45917 \\ \hline
\multicolumn{1}{|c|}{\textbf{10}} & 2.19788 & 2.5042 & 1.75817 & 1.92655 & 1.51911 & 1.94492 & 1.47031 & 1.54005 & 1.52353 & 1.76126 \\ \hline
\multicolumn{1}{|c|}{\textbf{15}} & 1.99967 & 2.39172 & 1.59689 & 1.96891 & 1.59307 & 1.68918 & 1.53418 & 1.50725 & 1.52251 & 1.53919 \\ \hline
\multicolumn{1}{|c|}{\textbf{20}} & 2.81062 & 2.92986 & 1.58119 & 2.40306 & 1.45906 & 1.76651 & 1.47680 & 1.59628 & 1.53945 & 1.51312 \\ \hline
\end{tabular}}
\end{table*}

\subsection{Case analysis of prediction quality of different target materials}

To further understand how the DeepGATGNN model performs for the vibration frequency prediction, we used our model trained with the Mixed dataset to predict 100 test samples and show results of six crystal structures with high prediction accuracy scores, including Fe$_2$H$_6$, B$_6$H18O18, B48O6, C44F28,  Be2BH3O5, C120F36. The six case study target materials contain binary, ternary and quaternary materials with diverse structures. The numbers of atoms within their unit cells range from 8 to 156. 

\begin{figure*}[htb!] 
    \begin{subfigure}[t]{0.33\textwidth}
        \includegraphics[width=\textwidth]{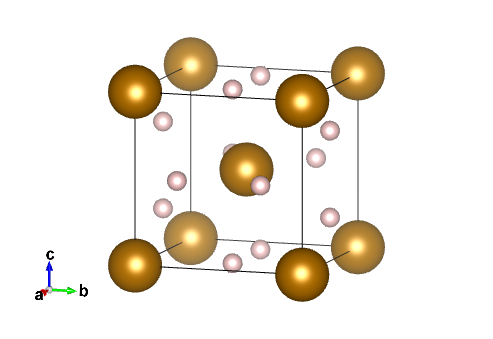}
        \caption{Fe$_2$H$_6$}
        \vspace{-3pt}
        \label{fig:Ce2As2O6_predim}
    \end{subfigure}\hfill
    \begin{subfigure}[t]{0.33\textwidth}
        \includegraphics[width=\textwidth]{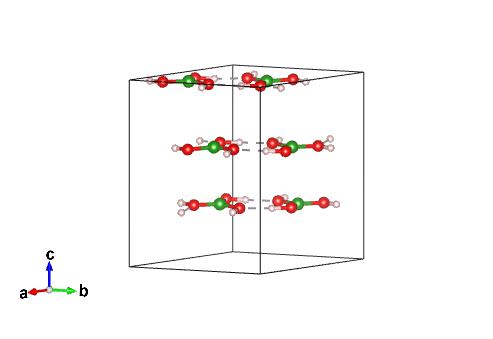}
        \caption{B$_6$H$_{18}$O$_{18}$}
        \vspace{-3pt}
        \label{fig:special_coordinates}
    \end{subfigure}\hfill
    \begin{subfigure}[t]{0.33\textwidth}
        \includegraphics[width=\textwidth]{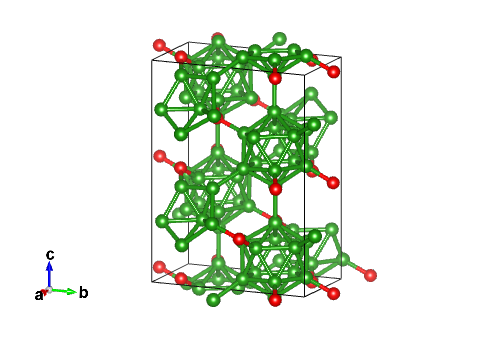} 
        \caption{B$_{48}$O$_6$}
        \vspace{-3pt}
        \label{fig:Ce2As2O6_predim}
    \end{subfigure}\hfill
    \vspace{1pt}
    \begin{subfigure}[t]{0.33\textwidth}
        \includegraphics[width=\textwidth]{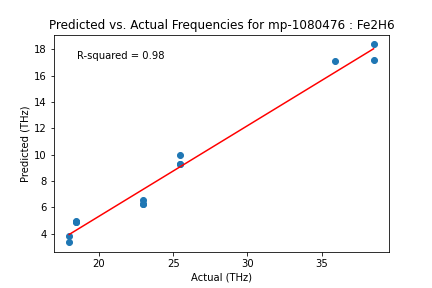}
        \caption{Fe$_2$H$_6$}
        \vspace{-3pt}
        \label{fig:special_coordinates}
    \end{subfigure}\hfill    
    \begin{subfigure}[t]{0.33\textwidth}
        \includegraphics[width=\textwidth]{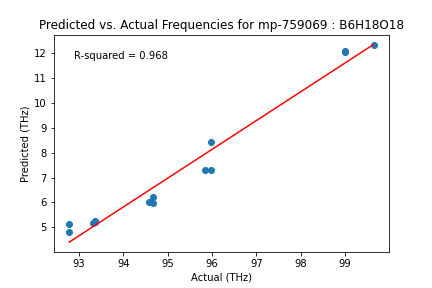}
        \caption{B$_6$H$_{18}$O$_{18}$}
        \vspace{-3pt}
        \label{fig:Ce2As2O6_predim}
    \end{subfigure}\hfill
    \begin{subfigure}[t]{0.33\textwidth}
        \includegraphics[width=\textwidth]{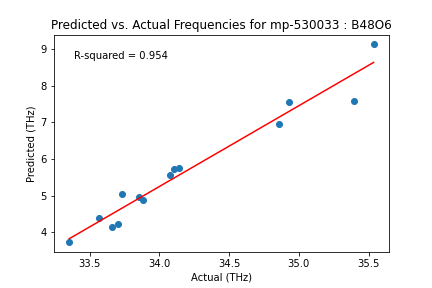}
        \caption{B$_{48}$O$_6$}
        \vspace{-3pt}
        \label{fig:special_coordinates}
    \end{subfigure}\hfill

    \begin{subfigure}[t]{0.33\textwidth}
        \includegraphics[width=\textwidth]{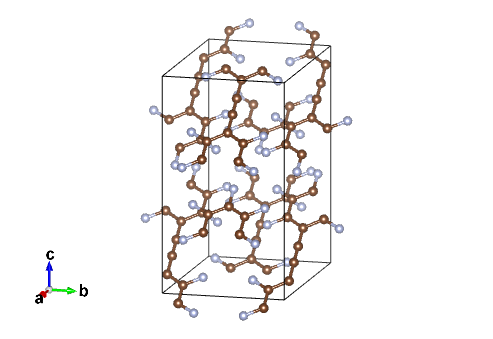}
        \caption{C$_{44}$F$_{28}$}
        \vspace{-3pt}
        \label{fig:j}
    \end{subfigure}\hfill
    \begin{subfigure}[t]{0.33\textwidth}
        \includegraphics[width=\textwidth]{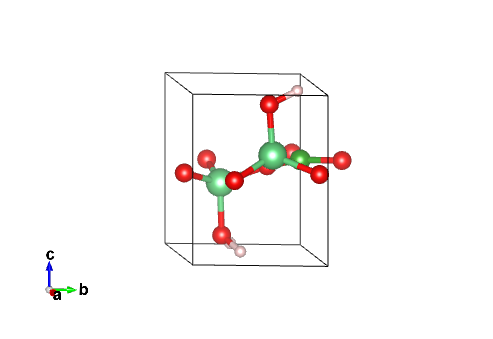}
        \caption{Be$_2$BH$_3$O$_5$}
        \vspace{-3pt}
        \label{fig:k}
    \end{subfigure}\hfill
    \begin{subfigure}[t]{0.33\textwidth}
        \includegraphics[width=\textwidth]{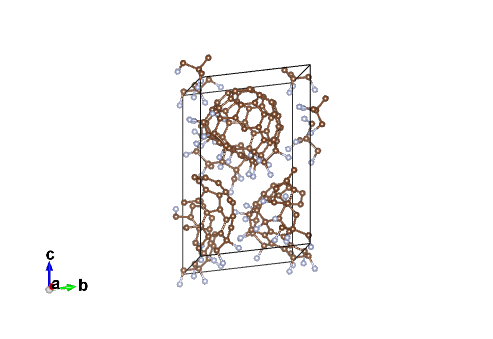}
        \caption{C$_{120}$F$_{36}$}
        \vspace{-3pt}
        \label{fig:l}
    \end{subfigure}\hfill
    \begin{subfigure}[t]{0.33\textwidth}
        \includegraphics[width=\textwidth]{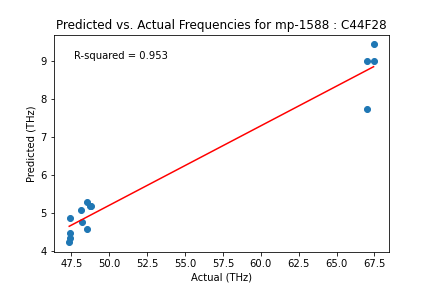}
        \caption{C$_{44}$F$_{28}$}
        \vspace{-3pt}
        \label{fig:g}
    \end{subfigure}\hfill    
    \begin{subfigure}[t]{0.33\textwidth}
        \includegraphics[width=\textwidth]{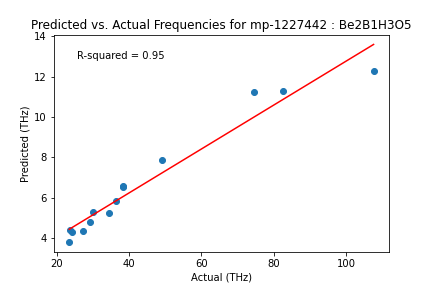}
        \caption{Be$_2$BH$_3$O$_5$}
        \vspace{-3pt}
        \label{fig:h}
    \end{subfigure}\hfill
    \begin{subfigure}[t]{0.33\textwidth}
        \includegraphics[width=\textwidth]{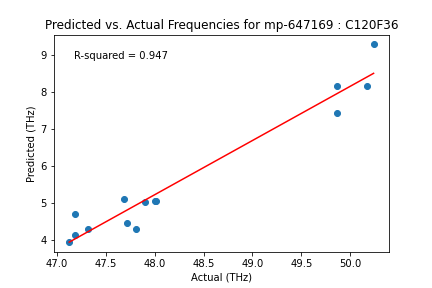}
        \caption{C$_{120}$F$_{36}$}
        \vspace{-3pt}
        \label{fig:i}
    \end{subfigure}\hfill

    \caption{Prediction performance of vibration frequencies by deeperGATGNN. Group one: (a)(b)(c)(h), structures of four materials Fe$_2$H$_6$, B$_6$H$_{18}$O$_{18}$, B$_4$8O6, Be2BH3O5 along with their predicted vibration frequencies in (d)(e)(f)(k) and the regression  $r^2$ scores of 0.98, 0.968, 0.954, 0.95 respectively. The vibration frequencies of this group are spread all over the whole range. Group two: (g)(i) show the structures of two materials C$_{44}$F$_2$, C$_{120}$F$_3$.  and their predicted frequencies in (j)(l) with $R^2$ scores of 0.953 and 0.947. Their vibration frequencies are clustered at two ends of the frequency range.}
    
    \label{fig:case_studies_performance}
\end{figure*}

In Figure \ref{fig:case_studies_performance}, we present each of the target structures and their scatter plots showing the predicted vibration frequencies versus the ground truths. 
We can divide them into two groups for discussion based on the distribution of their vibration frequencies. In group one, the frequencies are coarsely distributed evenly within the whole range of their vibration frequencies as shown in Figure \ref{fig:case_studies_performance}(d)(e)(f)(k). This group includes Fe2H6, B6H18O18, B48O6, Be2BH. For this group of materials, our DeepGATGNN model achieves very good performance with the $R^2$ scores of 0.98, 0.968, 0.954, 0.95 respectively. In group two, the vibration frequencies are distributed within two extreme clusters at the two ends of the frequency range as shown in Figure \ref{fig:case_studies_performance}(j)(l). It includes two materials: C44F28 and C120F30. Usually this type of distributions are difficult to achieve good regression results. However, our prediction model obtains high regression coefficient $R^2$ scores of  0.953 and 0.947 for C44F28 and C120F30 respectively. Overall, we find the $R^2$ scores are all above 0.9 for all six target structures: the best score is 0.98 for Fe2H6 and the lowest one is 0.947 for C120F36.

\section{Conclusion}

We have proposed a deep global graph attention neural network algorithm for the prediction of vibration frequency of a given crystal material given their structure information. We formulate it as a variable-dimension vector target regression problem. Extensive experiments on two datasets with 35,552 and 15,000 samples show that our graph network model can handle the varying sizes of the training samples and can predict the vibration frequency with good performance for the rhombohedral crystal materials with $R^2$ score reaching 0.724. For the dataset with mixed structures, the vibration frequency prediction is much more challenging with the $R^2$ score around 0.556. We find the increasing the number of training samples can significantly reduce the prediction error which is widely recognized in other materials property prediction tasks. However, our current model does make mistakes for some vibration frequencies by predicting them as zero. Further research is especially needed to build more accurate models trained with mixed structure types and used to predict the vibration frequency of diverse crystal structures.

\section{Conflict of interest}
The authors declare no competing financial interest.

\section{Contribution}
Conceptualization, J.H., M.H.; methodology, J.H., N.N., S.L., M.H.; software, N.N., S.L.; validation, N.N., M.H., J.H.; investigation, N.N., J.H., L.W., M.H.; writing–original draft preparation, J.H., L.W.,K.C., N.N., M.H.; writing–review and editing, J.H, L.W., M.H.; visualization, N.N.,M.H.; supervision, J.H.,M.H.; funding acquisition, J.H,M.H.

\section{Acknowledgement}
This research was funded by NASA SC Space Grant Consortium REAP Program (Award No.: 521179-RP-SC005). Research reported in this work was supported in part by NSF under grant and 1940099, 1905775, 2030128, and by NSF SC EPSCoR Program under award number (NSF Award OIA-1655740 and GEAR-CRP 19-GC02). This project is also partially supported by DOE under grant DE-SC0020272. The views, perspective, and content do not necessarily represent the official views of the SC EPSCoR Program nor those of the NSF.

\bibliographystyle{apsrev4-1}

\bibliography{references}  %

\end{document}